\documentclass{npw}
\usepackage{graphicx,color}

\def\d{\hbox{d}}
\def\be{\begin{equation}}
\def\ee{\end{equation}}
\def\bea{\begin{eqnarray}}
\def\eea{\end{eqnarray}}
\def\l{\label}
\def\r{{\bf r}}

\def\om{\omega}

\def\hahat{\hat{H}}

\def\hahat0{\hat{H}_0}

\def\cos{\hbox{cos}}
\def\sin{\hbox{sin}}

\def\d{\hbox{d}}
\def\eps{\epsilon}
\def\epsd{\varepsilon}

\def\epsi{{\cal E}}
\def\siml{\hbox{\kern.1em \lower.6ex \hbox{$\sim$} \kern-1.12em
 \raise.6ex \hbox{$<$} \kern.1em}}
\def\simg{\hbox{\kern.1em \lower.6ex \hbox{$\sim$} \kern-1.12em
 \raise.6ex \hbox{$>$} \kern.1em}}

\begin{document}

\title{ Derivative corrections to the symmetry energy and the 
isovector dipole-resonance structure in nuclei}
\author{
  J P Blocki \\
  {\it National Centre of Nuclear Research, PL-00681 Warsaw, Poland}\\
  A G Magner\email{magner@kinr.kiev.ua} \\
  {\it Institute for Nuclear Research,  Kyiv 03680, Ukraine}\\
P Ring\\
{\it Technical Munich University,  D-85747 Garching, Germany}
}
\date{}
\maketitle

\vspace*{-1.05cm}

\begin{abstract}
The effective surface approximation is extended accounting for  
derivatives of the symmetry energy density per particle.
Using the analytical isovector surface energy constants 
within the Fermi-liquid droplet model, one obtains 
energies and  
sum rules of the isovector 
dipole resonance structure in a reasonable agreement with  
the experimental data and other theoretical approaches.

\end{abstract}

 \section{Introduction}

The symmetry energy is a key quantity for study of the fundamental 
properties of exotic nuclei 
with a large excess of neutrons above protons in the nuclear and astronomic 
physics 
\cite{myswann,myswnp80pr,myswiat85,myswnp96pr,vinas1,vinas2,vinas3,vinas4}. 
In spite of a very intensive study of
these properties, the  derivatives
of the symmetry energy and its surface coefficient are still rather
undetermined 
in the calculations by the liquid droplet model (LDM) 
\cite{myswann}, 
or within more general  local density approach (LDA) \cite{gross1,gross2}, 
in particular 
within the extended Thomas-Fermi (ETF) approximation \cite{brguehak},  
and models based on the Hartree-Fock (HF) method \cite{brink}
with applying the Skyrme forces 
\cite{bender,chaban_npa,reinhardSV,pastore}, 
in contrast to the basic volume symmetry energy constant.
Within the nuclear LDA,
the variational condition derived from minimizing the nuclear energy 
at the fixed particle number and the neutron excess above protons 
can be simplified 
using the expansion
in a small parameter  $a/R\sim A^{-1/3}$ for heavy enough nuclei
with $a$ being of the order of
the diffuse edge thickness of the nucleus, 
$R$ the mean curvature radius of the ES, and
$A$ the number of nucleons within the effective surface
(ES) approximation 
\cite{strtyap,strmagbr,strmagden,magsangzh,BMRV} . A separation
of the nuclear energy into the volume and surface (and curvature) 
components of the LDM and ETF makes obviously sense for  $a/R \ll 1$.
The accuracy of the ES  approximation  
in the ETF approach \cite{brguehak} was checked 
\cite{strmagden} 
with the spin-orbit (SO) and asymmetry terms  
\cite{magsangzh,BMRV}
by comparing results with those of the Hartree-Fock (HF) and other 
ETF theories for some Skyrme forces.

Solutions for the isoscalar and isovector particle densities in the ES
approximation of the ETF approach were applied to the analytical 
calculations of the surface symmetry-energy constants and 
the neutron skin in the leading order of the parameter $a/R$ 
\cite{magsangzh,BMRV}. Our results are compared with the 
previous ones \cite{myswann,myswnp80pr,myswiat85,myswnp96pr} 
in the LDM and recent works 
\cite{vinas1,vinas2,vinas3,vinas4,agrawal,kievPygmy_voinov,kievPygmy_larsen,nester1,nester2}.

The structure of the isovector dipole-resonance (IVDR) strength 
in terms of the main and satellite (pygmy) modes
\cite{kievPygmy_voinov,kievPygmy_larsen,nester1,nester2,dang1998,dang2001,ponomarev,adrich,wieland,nakatsukasa2012,BMRPhysScr2014}  
as a function of the isovector surface 
energy constant in the ES approach
can be described within the  Fermi-liquid droplet (FLD) 
model \cite{kolmag,kolmagsh}. 
The analytical expressions for the surface symmetry
energy constants are tested by the energies and sum rules 
of the isovector 
dipole resonance (IVDR) strength structure 
within the FLD model \cite{BMRPhysScr2014}.

In the present work, we shall extend the variational effective surface
method \cite{BMRV,BMRPhysScr2014} taking into account also derivatives of 
the non-gradient 
terms in the symmetry energy density along with its 
gradient ones. The fundamental isovector derivative and 
surface-tension constants are not
fixed yet well enough by using the experimental data for the neutron
skin thickness \cite{vinas4}. We suggest to use also the empirical data 
for the splitting of the IVDR strength structure to evaluate them better 
through their
analytical ETF expressions in the ES approximation 
as functions of the Skyrme force parameters. 

\section{Symmetry energy and particle densities}

We start with the nuclear energy as a functional of 
the isoscalar and isovector particle densities
$\rho_{\pm}=\rho_n \pm \rho_p$ in the local density approach
\cite{brguehak,bender}:
$E=\int \d \r\; \rho_{+}\epsi\left(\rho_{+},\rho_{-}\right)$, where
\bea\label{enerden}
&&\epsi\left(\rho_{+},\rho_{-}\right) \approx
- b^{}_V 
+ J I^2 + \epsd_{+}(\rho_{+},\eps) +
\epsd_{-}(\rho_{+},\rho_{-},\eps) +
\nonumber\\
&+&
\left({\cal C}_{+}/\rho_{+} +{\cal D}_{+} \right)
\left(\nabla \rho_{+}\right)^2
+ \left({\cal C}_{-}/\rho_{+} + {\cal D}_{-}\right)
\left(\nabla \rho_{-}\right)^2
\eea
and
\bea\l{surfsymen}
\epsd_{-}(\rho_{+},\rho_{-},\eps)&=&
\mathcal{S}_{\rm sym}(\eps)
\left(\rho_{-}/\rho_{+}\right)^2 - J I^2,\nonumber\\
\mathcal{S}_{\rm sym}(\eps)&=&J - L \eps + K_{-}\eps^2/2.
\eea
We introduced also a small parameter  $\eps$
of the expansion following suggestions of
Myers and Swiatecki \cite{myswann},
$\eps=(\rho_\infty - \rho_{+})/(3\rho_\infty)$, 
where $\rho_\infty=3/4\pi r_0^3 \approx$ 0.16 fm$^{-3}$
is the density of the infinite nuclear matter and $r_0 = R/ A^{1/3}$ 
is the radius constant. In (\ref{enerden}),
$b^{}_V \approx$ 16 MeV is the separation energy per particle.
The isoscalar part
of the surface energy-density (\ref{enerden}) (zero-order terms in
expansion over $\rho_{-}/\rho_{+}$, denoted as $\delta$  
\cite{myswann}) which does not depend explicitly
on the density gradient terms, is determined by the function
$\epsd_{+}(\rho_{+},\eps)$. 
We use a representation 
$\epsd_{+}=K_{+} 
\overline{\epsd}_{+}/18$ by the dimensionless quantity 
$\overline{\epsd}_{+}$.
$K_{+}$ is the isoscalar 
in-compressibility modulus of the symmetric nuclear matter, 
$K_{+} \approx 220-245$ MeV, 
\be\l{epsilonplus}
\overline{\eps}_{+}(\rho_{+},\eps)=9 \eps^2 + I^2 
\left[\mathcal{S}_{\rm sym}(\eps) -J\right]
/K_{+},
\ee
$~I=(N-Z)/A$ is the asymmetry parameter,
$~N=\int \d \r \rho_n(\r)$ and $~Z=\int \d \r \rho_p(\r)$ are
the neutron and proton numbers and $~A=N+Z$. 
The second term, $\propto I^2$, appears
due to the derivative corrections to the volume symmetry 
energy. The symmetry energy constant of the nuclear matter $~J \approx$
30 MeV specifies the main volume term in the symmetry energy.
In contrast to the simplest approximation \cite{BMRV}, terms with
$L \approx 20 \div 120$ MeV, and even less known 
$K_{-} \approx -470 \div 140$ MeV \cite{vinas1,vinas4}, defined by the first 
and second derivatives of the symmetry energy expansion 
with respect to the variable 
$\eps$ were taken into account (\ref{surfsymen}), (\ref{epsilonplus}).   
Equation (\ref{enerden})
can be applied in a semiclassical approximation for realistic
Skyrme forces \cite{chaban_npa,reinhardSV,pastore},
in particular by neglecting higher $\hbar$ corrections 
(the ETF kinetic energy) \cite{brguehak}
and Coulomb terms \cite{strmagden,magsangzh,BMRV}. 
Up to a small Coulomb exchange terms they all can be easily taken into account  
\cite{strtyap,BMRV}. Constants ${\cal C}_{\pm}$ and ${\cal D}_{\pm}$ 
are defined by parameters of the Skyrme forces
 \cite{brguehak,bender,chaban_npa,reinhardSV,pastore}. For ${\cal C}_{\pm}$
one has
\bea\l{Cpm}
{\cal C}_{+}&=& 
\frac{1}{12} \left(t_1 - \frac{25}{12} t_2 - \frac{5}{3} t_2 x_2\right),
\nonumber\\
{\cal C}_{-}&=&-\frac{1}{48} t_1\left(1 + \frac{5}{2}x_1\right) - 
\frac{1}{36} t_2\left(1 + \frac{19}{8} x_2\right).
\eea
The isoscalar SO gradient terms in (\ref{enerden}) are defined 
with a constant: ${\cal D}_{+} = -9m W_0^2/16 \hbar^2$, where
$W_0 \approx$100 - 130~ MeV$\cdot$fm$^{5}$ and
$m$ is the nucleon mass \cite{brguehak,bender,chaban_npa,reinhardSV}.
The isovector SO constant 
${\cal D}_{-}$ is usually relatively small and will be neglected here
for simplicity. We emphasize that (\ref{enerden}) has a general form 
\cite{strmagbr,strmagden,magsangzh} for any densed system with a sharp edge.
In particular, it can be derived from 
comparison with one of energy density functionals determined by a 
Skyrme force in order to get  
relations (\ref{Cpm}) for ${\cal C}_{\pm}$ and other ones in terms of 
its parameters.

Equation (\ref{enerden}) contains
the volume component given by the first two terms 
and the surface part including the $L$ and $K_{-}$  derivative corrections
of the $\epsd_{-}$  and density gradients which both are important
for the finite nuclear systems
 \cite{BMRV}. These gradient terms together with other surface terms
of the energy density in the ES approximation are responsible for
the surface tension in finite nuclei.

Minimizing the energy $E$ under constraints
of the fixed particle number $A=\int \d \r\; \rho_{+}(\r)$ and
 neutron excess $N-Z= \int \d \r\; \rho_{-}(\r)$
(also other constraints as a deformation, 
nuclear angular momentum and so on 
\cite{strtyap,strmagbr,strmagden,BMRV}),
one arrives at the Lagrange equations with the isoscalar and isovector 
chemical potentials as corresponding multipliers.
The analytical solutions will be obtained approximately 
up to the order $A^{2/3}$ in the
binding energy. To satisfy these constraints for calculations
of the particle densities, one needs the leading order terms
in $ a/R \sim A^{-1/3}$
for calculations of the particle densities $\rho_{\pm}$. 
Using these densities for the
nuclear energy calculations
with a required accuracy we account for higher (next) order 
surface corrections in $ a/R$ with respect to the leading
order terms \cite{BMRV}.

\section{Extended densities and energies}

For the isoscalar particle density, $w =w_{+}=\rho_{+}/\rho_\infty$, one has
up to the {\it leading} terms in the parameter  $a/R$ the usual 
first-order differential Lagrange equation \cite{strmagden,magsangzh,BMRV}
but with the
solution depending on $L$
and $K_{-} $ through $\epsd_{+}$ (\ref{epsilonplus}),
\be\l{ysolplus}
{}\hspace{-2.0ex}x=-
\int_{w_r}^{w}\d y \sqrt{\frac{1 +\beta y}{y\overline{\eps}_{+}(y,\eps)}},\quad
x=\frac{\xi}{a}, 
\ee
for $x < x(w=0)$ and $w=0$ for $x \geq x(w=0)$ where $x(w=0)$ 
is the turning point, $\xi$ is the distance from a given spatial 
point to the ES
in a local coordinate system $(\xi,\eta)$ ($\xi=r-R$ for spherical nuclei 
and $\eta$ is the tangent-to-ES coordinate  \cite{strmagbr}). 
$\overline{\eps}_{+}(y,\eps)$ (\ref{epsilonplus}) is the dimensionless
energy density $\epsd_{+}$ per particle (\ref{enerden}) and
$\beta={\cal D}_{+} \rho_\infty/{\cal C}_{+}$ is the dimensionless SO
parameter.  For $w_r=w(x=0)$
up to $I^2$ corrections, one has the boundary condition,
$\d^2 w(x)/\d x^2=0$ at the ES ($x=0$):
\begin{equation}
\overline{\eps}_{+}(w_r)+
w_r(1 +\beta w_r) \left[\d \overline{\eps}_{+}(w)/\d w\right]_{w=w_r}=0\;.
\label{boundcond}
\end{equation}
In Eq.\ (\ref{ysolplus}), $a \approx 0.5-0.6$ fm is the
diffuseness parameter \cite{BMRV},
\be\l{adif}
a=\sqrt{\frac{{\cal C}_{+} \rho_\infty K_{+}}{30 b_V^2}},
\ee

For the isovector density, $w_{-}(x)=\rho_{-}/(\rho_{\infty}I) $,
after simple transformations of the isovector Lagrange equation
up to the leading term in $a/R$  in the ES approximation, one similarly
finds the
equation and boundary condition for $w_{-}(w)$ \cite{BMRV},
\be\l{yeq0minus}
\frac{\d w_{-}}{\d w} =c_{\rm sym}
\sqrt{\frac{\mathcal{S}_{\rm sym}(\eps)(1+ \beta w)}{J \overline{\eps}_{+}(w)}}
\;\sqrt{\Big|1 -\frac{w_{-}^2}{w^2}\Big|},
\ee
and $ w_{-}(w=1) =1$; $\mathcal{S}_{\rm sym}(\eps)$ is given by 
(\ref{surfsymen}) and
\be\l{csym}
c_{\rm sym}=a \left[J/(\rho_\infty
\vert{\cal C}_{-}\vert)\right]^{1/2},
\ee
see \cite{BMRV} for the case  of the constant $\mathcal{S}_{\rm sym}=J$.
The analytical solution  $w_{-}=w \cos[\psi(w)]$
can be obtained through the expansion 
of $\psi$ in powers of
\be\l{gamma}
\gamma(w)=3\eps/c_{\rm sym}.
\ee
Expanding up to the second order in $\gamma$, one finds
\cite{BMR2015}
\be\l{ysolminus}
w_{-} =  w\;\cos\psi(w) \approx 
w \left(1- \psi^2(w)/2 + ...
\right),
\ee
where
\be\l{psi2}
\psi(w)=\gamma(w) 
\left[1 + \widetilde{c} \gamma(w) + ...\right]/\sqrt{1+\beta},
\ee
\be\l{ctilde}
\widetilde{c}=[\beta c_{\rm sym}^2 + 2 + 
c_{\rm sym}^2 L (1+\beta)]/[3 (1+\beta)].
\ee
The $L$ dependence of $w_{-}$ appears
at the second order
in $\gamma$ but $w_{-}$ is independent of $K_{-}$ at this order.
Note that $K_{-}$ is the coefficient in expansions (\ref{surfsymen}) and 
(\ref{epsilonplus}) at higher order $\eps^2$, and therefore,
it shows up at higher (third) 
order terms of the  expansion (\ref{ysolminus}) in $\gamma$  
\cite{BMRV}. 

\begin{figure}
\begin{center}
\includegraphics*[width=0.45\textwidth]{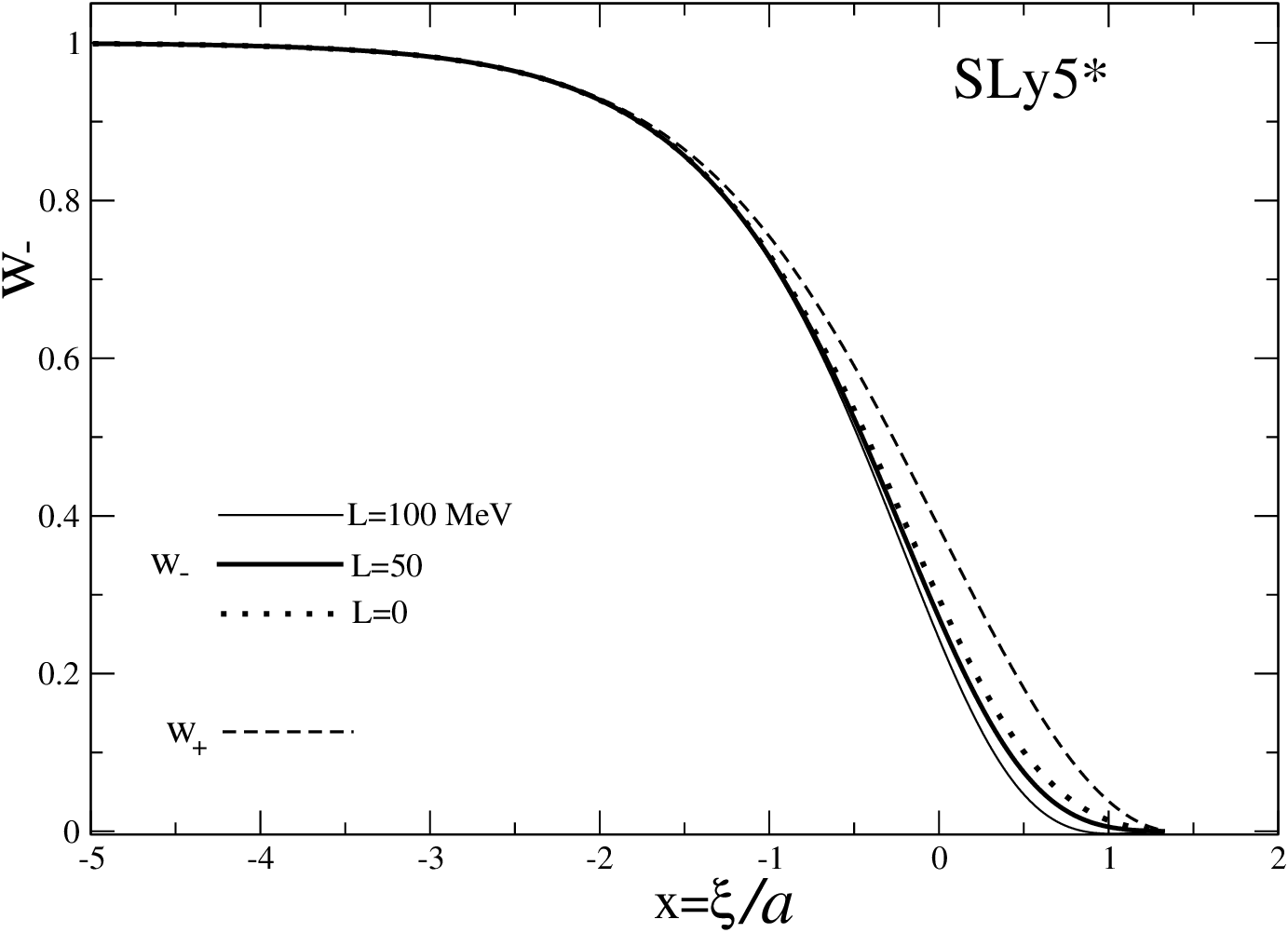}
\end{center}

\vspace{-0.8cm}
\caption{Isovector $w_{-}$
(\ref{ysolminus}) [with ($L=50$ and $100$ MeV)
and without ($L=0$) derivative constant $L$] and isoscalar $w=w_{+}$ 
(see \cite{BMRV}) particle densities are shown 
vs $x=\xi/a $ 
for the Skyrme force SLy5$^*$ ($x\approx (r-R)/a$ for small nuclear 
deformations \cite{pastore,BMRPhysScr2014}).
}
\label{fig1}
\end{figure}

\begin{figure}
\begin{center}
\includegraphics*[width=0.45\textwidth]{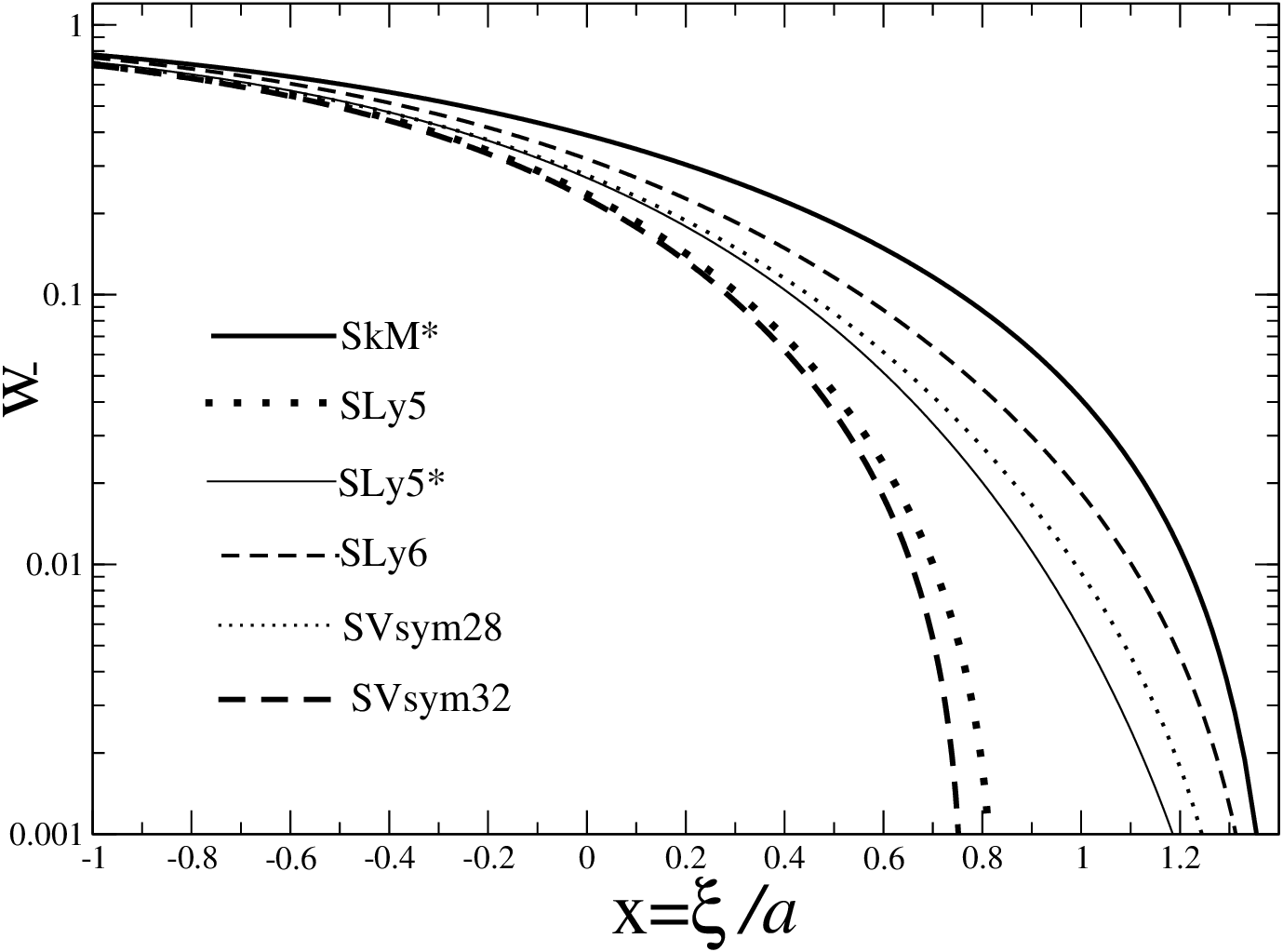}
\end{center}

\vspace{-0.8cm}
\caption{
Isovector density
$w_{-}(x)$ (\ref{ysolminus}) (in the logarithmic scale)  as function of $x$ 
within the quadratic approximation to
$\overline{\epsd}_{+}(w)$ 
for several Skyrme forces \cite{chaban_npa,reinhardSV,pastore}; 
($L=48.269$ MeV for SLy5 and
$L=47.449 $ MeV for SLy6 \cite{jmeyer}, which are approximately  
$50$ MeV within a precision
of the line thickness; and a similar approximation was used for other forces.
}
\label{fig2}
\end{figure}

In Fig.\ \ref{fig1}, the $L$ dependence of the 
function $w_{-}(x)$ within rather a large interval $L=0-100$ MeV 
\cite{vinas2} is shown as compared to
that of the density $w(x)$ for the SLy5$^*$ force as a typical example
($L=45.885 \approx 50$ MeV \cite{jmeyer} and Table I).
As  shown in Fig.\ \ref{fig2} in a 
logarithmic scale,
one observes a big difference in the isovector densities $w_{-}$ derived 
from different Skyrme forces \cite{chaban_npa,reinhardSV,pastore}
within the diffuse edge. All these calculations have been done 
with the finite
 value of the slope parameter $L$ taken approximately from 
\cite{reinhardSV,jmeyer} (Table I), which 
is important for calculations of the neutron skins of nuclei.
The isovector 
particle density $w_{-}$ (\ref{ysolminus}) in the second order of the
small parameter $\gamma$ does not depend on the symmetry energy 
in-compressibility $K_{-}$ as mentioned above.  
Therefore, it is possible to study first the main slope effects of 
$L$ neglecting small $I^2$ corrections to the
isoscalar particle density $w_{+}$ \cite{BMRV} through the 
$\overline{\epsd}_{+}$ (\ref{epsilonplus}). 
Then, we may deal with more precisely the effect of the second derivatives 
$K_{-}$ taking into account higher order terms.

We emphasize that the dimensionless densities, $w(x)$ 
\cite{BMRV} and $w_{-}(x)$ 
(\ref{ysolminus}), shown in Figs.\ \ref{fig1} and \ref{fig2} were 
obtained in 
the leading ES approximation ($a/R \ll 1$) as functions of the 
specific combinations
of the Skyrme force parameters like $\beta$, $ c_{\rm sym}$ \cite{BMRV}  
but taking into account now the $L$-dependence.  
They are the universal distributions
independent of the specific properties of the nucleus \cite{BMRV}.
It yields approximately the local density distributions in the 
normal-to-ES direction $\xi$  
with the correct asymptotical behavior outside of the ES layer for any 
deformation  
at $a/R \ll 1$, 
as well as for the semi-infinite nuclear matter. 
Densities $w_{\pm}$
are universal distributions independent of the specific properties of nuclei.

The nuclear energy $E$
in this improved ES approximation 
is split into the volume and
surface 
terms \cite{BMRV},
\be\l{EvEs}
E \approx -b^{}_V\; A + J (N-Z)^2/A + E_S^{(+)} + E_S^{(-)}.
\ee
For the isoscalar (+) and isovector (-) surface energy
 components $E_S^{(\pm)}$, one obtains
 $E_S^{(\pm)}= b_S^{(\pm)} S/(4\pi r_0^2)$,
where $S$ is the surface area of the ES, $b_{S}^{(\pm)}$
are the
isoscalar $(+)$ and isovector $(-)$ surface energy constants,
\be\l{sigma}
b_S^{(\pm)} \approx
8 \pi r_0^2 {\cal C}_{\pm}\int_{-\infty}^{\infty} \d \xi\;
\left(1 + {\cal D}_{\pm}\rho_{+}/{\cal C}_{\pm}\right)
\left(\partial \rho_{\pm}/\partial \xi\right)^2.
\ee
These constants are proportional to the corresponding surface tension 
coefficients
$\sigma_{\pm} = b_S^{(\pm)}/(4 \pi r_0^2)$ through 
solutions 
for $\rho_{\pm}(\xi)$ [see (\ref{ysolplus}) and \ref{ysolminus})] 
which can be taken into account in the leading order 
of $a/R$ .  
These
coefficients $\sigma_{\pm}$ are the same as found in 
expressions for the capillary pressures of the macroscopic 
boundary conditions [see \cite{BMRV} extended to new $\epsd_{\pm}$ 
(\ref{surfsymen}) and (\ref{epsilonplus}) modified
by $L$ and $K_{-}$ derivative corrections].
Within the improved ES approximation where 
higher order corrections in the small parameter $a/R$ 
are taken into account, we derived equations for the
nuclear surface itself. 
For more exact isoscalar and isovector particle densities we account for the
main terms in the next order of the parameter $a/R$
in the Lagrange equations \cite{BMR2015}.
Multiplying these equations by $\partial \rho_{-}/\partial \xi$
and integrating them over the ES
in the  
direction $\xi$ and using
solutions for $w_{\pm}(x)$ up to the leading orders
[see (\ref{ysolplus}) and
(\ref{ysolminus})],
one arrives at the ES equations in the form of
the macroscopic boundary conditions
\cite{BMRV,kolmagsh}.
They ensure equilibrium through the equivalence of the volume and 
surface (capillary) pressure
(isoscalar or isovector) variations. 
The latter ones are proportional to the 
surface tension coefficients $\sigma_{\pm}$.
For the energy surface coefficients $b_{S}^{(\pm)}$ (\ref{sigma}), one obtains
\be\l{bsplus}
{}\hspace{-2ex}b_S^{(+)}\!= \!6{\cal C}_{+}
\frac{\rho_\infty {\cal J}_{+}}{ r_0 a},\,\,\,
{\cal J}_{+}\!=\!\int_0^1 \d w
[w (1+\beta w) \overline{\epsd}_{+}(w)]^{1/2},
\ee
 and 
\be\l{bsminus}
b_S^{(-)}=k^{}_S\; I^2,\qquad
k^{}_S=6 {\cal C}_{-} \frac{\rho_\infty {\cal J}_{-}}{r_0 a},
\ee
where
\bea\l{jminusdef}
{\cal J}_{-}&=&\int_0^1 \!\d w\left[
\frac{w \overline{\eps}_{+}(w)}{(1\!+\!\beta w)}\right]^{1/2}
\left\{\!\cos\psi \right.\nonumber\\
&+&\left. 
\left[w\sin\psi/\left(c_{\rm sym}
\sqrt{1\!+\!\beta}\right)\right]
\left[1\!+\!2 \widetilde{c} \gamma(w)\right]\right\}^2. 
\eea
For $\gamma$ and $\widetilde{c}$, 
see (\ref{gamma}) and (\ref{ctilde}).  For ${\cal J}_{-}$ 
one can use the following approximation:
\bea\l{jminus}
{\cal J}_{-}
&\approx&\int_0^1 (1-w) \d w\;
\sqrt{\frac{w}{1+\beta w}}\;
\left\{1+\frac{2\gamma(w)}{c_{\rm sym}(1+\beta)} \right.\nonumber\\ 
&+&{}\hspace{-0.5cm}
\;\left.\frac{\gamma^2}{(1+\beta)^2}\left[\frac{1}{c_{\rm sym}^2} +
6 (1+\beta)\left(\frac{\widetilde{c}}{c_{\rm sym}} -
\frac12\right)\right]\right\}.
\eea
Simple expressions for constants $b_S^{(\pm)}$ 
can be easily derived  in terms of the algebraic and trigonometric functions
for the quadratic form of $\overline{\eps}_{+}(w,\eps)$ ($\eps=0$). 
In these derivations we neglected curvature terms and, being of the same order,
shell corrections. The isovector energy terms were obtained within the ES
approximation with the high accuracy up to a small $I^2 (a/R)^2$.

According to the macroscopic 
theory 
\cite{myswann,strtyap,strmagbr,strmagden,magsangzh,BMRV},
one may define the isovector stiffness $Q$ with respect to the difference
$R_n-R_p$ 
between the neutron and proton radii as a collective variable,
\be\l{stifminus}
Q=-k^{}_S I^2/\tau^2,
\ee
where $\tau$ is the neutron skin in $r_0$ units.
Defining the neutron and proton radii $R_{n,p}$ as the positions of the
maxima of the neutron and proton density gradients, respectively, one obtains
the neutron skin $\tau$ \cite{BMRV}, 
\be\l{skin}
\tau= \frac{R_n-R_p}{r_0} \approx 
\frac{8 a I g(w_r)}{r_0 c_{sym}^2}, 
\ee
with
\bea\l{fw}
&&g(w)=\frac{w^{3/2}(1+\beta w)^{5/2}}{
(1+\beta)(3w+1+4\beta w)} \left\{w
\left(1+2\widetilde{c} \gamma\right)^2 +\right.\nonumber\\ 
&&\left.
2\gamma  \left(1+\widetilde{c}
\gamma\right)\left[\widetilde{c}
w - c_{sym}\left(1+
2 \widetilde{c} \gamma\right)\right]\right\},
\eea
taken at the ES value $w_{r}$ [$w''(w_r)=0$ (\ref{boundcond})].
Accounting also for 
(\ref{bsminus}) and (\ref{jminus}),
one finally arrives at
\be\l{stiffin}
Q=-\nu\; J^2/k^{}_S,  \quad
\nu=
9 {\cal J}_{-}^2/[16 g^2(w_r)],
\ee
where ${\cal J}_{-}$ and $g(w)$ are given by (\ref{bsminus}), (\ref{jminus})
and (\ref{fw}), respectively.
This $Q$ with $\nu=9/4$ has been predicted earlier
\cite{myswann}.
However, in our derivations $\nu$ deviates from $9/4$, and it is 
proportional to the function
${\cal J}_{-}^2/g^2(w_r)$. This function depends 
significantly on the SO interaction
parameter $\beta$ but not too much on the specific Skyrme forces 
\cite{BMRV,BMRPhysScr2014}. The isovector stiffness coefficient 
$Q$ (\ref{stiffin}) depend essentially on constants 
$\mathcal{C}_{-}$ and $L$ (also $K_{-}$) through $\nu$ (\ref{stiffin})
and $k^{}_S$ (\ref{bsminus}) (and equations (\ref{epsilonplus}) 
and (\ref{surfsymen})).

The universal functions $w(x)$ 
\cite{BMRV} and $w_{-}(x)$ 
(\ref{ysolminus}) in the leading order of the ES approximation
can be used [analytically in the quadratic
approximation for $\overline{\epsd}_{+}(w)$] for 
calculations of the surface 
energy coefficients
$b_{S}^{(\pm)}$ (\ref{sigma}) and the neutron skin $\tau \propto I$.
As it was shown 
\cite{BMRV},   
only these particle 
density distributions  $w(x)$  and $w_{-}(x)$ within the surface layer 
are needed through their derivatives.

Therefore,  the surface symmetry-energy coefficient 
$k_S$ (\ref{bsminus}),  
the neutron skin 
$\tau$ (\ref{skin}) 
and the isovector stiffness 
$Q$ (\ref{stiffin})
can be approximated analytically in terms of functions 
of the 
critical surface combinations of the Skyrme parameters  $a$, $\beta$,
$b^{(\pm)}_{S}$, $c_{\rm sym}$, 
as well as the volume ones
$\,\rho_\infty$, $K$, $J$, and derivatives of the symmetry energy 
$L$ and $K_{-}$. 
Thus,  they are independent of the specific properties 
of the nucleus 
in the ES approximation.  

%
\noindent
\hspace{-1.5cm}
\begin{table}[pt]
\begin{center}
\begin{tabular}{|c|c|c|c|c|}
\hline
  & SLy5 & SLy5$^*$ & SVsym28 & SVsym32 \\
\hline
$L$(MeV) & 50 & 50 & 10 & 60   \\
$k_{S,0}$(MeV) & -12.6 & -13.1 & 11.4 & 15.6   \\
$k^{}_S$(MeV) & -13.8 & -15.0 & 11.6 & 17.6  \\
$\nu_0$  & 0.37 & 0.92 &
   0.90  & 0.84 \\
$\nu$  & 0.66 & 0.60 &
  0.83  & 0.69 \\
$Q_0$ (MeV) & 73 & 72 & -62 & -55    \\
$Q$(MeV)  & 49 & 41 & -56 & -37   \\
$\tau_0/I$ & 0.54 & 0.43 & 0.43 & 0.53    \\
$\tau/I$ & 0.53 & 0.60 & 0.45 & 0.69   \\
$D_0$(MeV)  
&  101 & 89  & 78 & 79  \\
\hspace{-2ex}$D$(MeV) 
& 100  & 89  & 78 &  78    \\
\hline
\end{tabular}
\end{center}

\vspace{0.2cm}
\small{Table\ I. Isovector energy coefficients 
$k^{}_S$ and stiffness $Q$ (\ref{stiffin}) (in units MeV), factor $\nu$, 
neutron skin $\tau/I$, and constant $D=\hbar \om^{}_{FLD} A^{1/3}$ of the 
IVGDR excitation energy $\hbar \om^{}_{FLD}$ (in MeV) for the $^{132}$Sn
are shown for a few Skyrme forces at different slope parameters $L$ 
\cite{agrawal,reinhardSV,jmeyer};
the zero low index means that $L$ is neglected;
the relaxation time $T$ \cite{belyaev} is
the same as used in Figures.} 
\end{table}

\section{Discussion of the results}

In Table I 
we show the isovector energy coefficient $k^{}_S$ [Eq.\ (\ref{bsminus})],
the isovector stiffness parameter $Q$ 
(\ref{stiffin}), its constant $\nu$ 
and the neutron skin $\tau$ (\ref{skin}).  
They were obtained for a few Skyrme forces
\cite{chaban_npa,reinhardSV,pastore} with different values of $L$
 within the ES approximation using the quadratic approximation 
for $\overline{\epsd}_{+}(w,\eps=0)$ \cite{BMRV}, and
neglecting the $I^2$ slope 
corrections.  
We also show the quantities $k_{S,0}$, $\nu_0$, 
$Q_0$ and $\tau_0$
where the slope corrections are neglected ($L=0$). 
In contrast to a fairly good agreement for the analytical isoscalar
energy constant $b_S^{(+)}$ \cite{BMRV} with that of 
\cite{BMRV,chaban_npa,reinhardSV,pastore},
the isovector energy coefficients $Q$ (or $k^{}_S=\nu J^2/Q$) are more 
sensitive to the choice
of the  Skyrme forces than the isoscalar ones \cite{BMRV}. 
The modula of  $Q$ are significantly smaller for most of the 
Skyrme forces SLy \cite{chaban_npa,pastore} and SV \cite{reinhardSV} 
than for the other ones, in contrast to the symmetry energy constant $k^{}_S$
for which one has the opposite behavior. However,
the $L$ dependence of $k^{}_S$ is not more pronounced  than that of $Q$
(Table I and II). For SLy and SV forces
the stiffnesses $Q$ are correspondingly significantly smaller in absolute value 
being more close to the well-known empirical
values of $Q $ \cite{myswnp96pr}
and semi-microscopic HF calculations \cite{vinas2,brguehak}, especially
with increasing $L$. Note that the $|Q|$ is rather more decreased with $L$ than
$k^{}_{S}$, sometimes for SLy and SV in factor of about two. 
Thus, the isovector stiffness $Q$
is even much more sensitive to constants of the Skyrme force and 
to the slope parameter $L$ than  
constants $k^{}_S$.
\noindent
\hspace{-1.5cm}
\begin{table*}[pt]
\begin{center}
\begin{tabular}{|c|c|c|c|c|c|c|c|c|c|}
\hline
 Skyrmes & L & $E_1$ & $E_2$ & $S_1$ & $S_2$ & D & $k^{}_S$ & $Q$ & $\tau/I$ \\
        & MeV & MeV & MeV & \% & \% & MeV & MeV & MeV &  \\
\hline
SLy5$^*$ $^{132}$Sn &  0 & 17.15 & 19.85 & 87.9 & 12.1 
        & 88.9 & -13.1 & 72 & 0.43 \\
         & 50 & 17.18 & 19.86 & 88.8 & 11.2 & 88.8 & -14.2 & 48 & 0.54 \\
         & 60 & 17.18 & 19.86 & 89.0 & 11.0 & 88.8 & -14.5 & 45 & 0.57\\
         & 70 & 17.18 & 19.87 & 89.2 & 10.8 & 88.7 & -14.8 & 42 & 0.59\\
         & 80 & 17.19 & 19.87 & 89.5 & 10.5 & 88.8 & -15.1 & 40 & 0.62\\
         & 90 & 17.19 & 19.87 & 89.7 & 10.3 & 88.8 & -15.5 & 38 & 0.64\\
         &100 & 17.20 & 19.87 & 89.9 & 10.1 & 88.8 & -15.8 & 36 & 0.67  \\
{}\hspace{7.0ex} $^{208}$Pb & 50 & 14.65 & 17.14 & 91.5 & 8.5 
      & 87.9 & -14.2 & 48 & 0.54\\
\hline
SVsym32 $^{132}$Sn &  0 & 14.70 & 18.00 & 72.9 & 27.1 & 78.8 
       &  15.6 &-55 & 0.53 \\
         & 50 & 14.69 & 17.99 & 75.1 & 24.9 & 78.4 &  17.2 &-39 & 0.66  \\
         & 60 & 14.69 & 17.98 & 75.5 & 24.5 & 78.3 &  17.6 &-37 & 0.69  \\
         & 70 & 14.69 & 17.98 & 76.0 & 24.0 & 78.2 &  18.0 &-35 & 0.71  \\
         & 80 & 14.69 & 17.98 & 76.4 & 23.6 & 78.2 &  18.4 &-33 & 0.74  \\
         & 90 & 14.68 & 17.97 & 76.9 & 23.1 & 78.3 &  18.9 &-32 & 0.77  \\
         &100 & 14.68 & 17.97 & 77.4 & 22.6 & 78.0 &  19.3 &-30 & 0.80\\
{}\hspace{7.0ex}$^{208}$Pb & 60 & 12.71 & 15.56 & 83.1 & 16.9 
      & 77.7 & 17.6 & -37 & 0.69\\
\hline
\end{tabular}
\end{center}

\vspace{0.2cm}
\small{Table\ II. Energies $E_n$ ($n=1,2$), EWSRs $S_n$, average 
IVDR (IVGDR) energy constants $D$, surface symmetry energy $k^{}_S$ and
isovector stiffness $Q$ constants, and neutron skin $\tau/I$ 
vs the slope parameter L in a typical region \cite{vinas4} for the same 
Skyrme forces SLy5$^*$ or SVsym32 and relaxation time $T$ as in 
Tables I and Figures; index 1 corresponds to the main mode
and index 2 to the satellite one; the results for $^{208}$Pb are 
shown for the comparison.} 
\end{table*}

Swiatecki and his collaborators \cite{myswnp96pr}
suggested the isovector stiffness values $Q\approx 30-35$ MeV 
accounting for the additional experimental data. 
More precise $A$-dependence of the averaged IVDR 
(Isovector Giant Dipole Resonance, IVGDR) energy $D_{FLD}$
for the finite values of $Q$ seems to be beyond the accuracy of
the both hydrodynamical and FLD model calculations. 
More realistic self-consistent  HF calculations with the Coulomb
interaction, surface-curvature and quantum-shell effects  
lead to larger
$Q\approx 30-80$ MeV \cite{vinas2,brguehak}.

The IVGDR energies and the energy weighted sum
rules (EWSR) obtained within
the semiclassical FLD approach based on the Landau-Vlasov equation
\cite{kolmagsh,BMRPhysScr2014} with the macroscopic boundary conditions 
\cite{BMRV} are also basically insensitive to the isovector surface energy
constants $k^{}_S$ or $Q$. They are similarly
in good agreement with the experimental data, and do not depend much
on the Skyrme forces, also when the finite slope symmetry energy
parameter $L$ is included 
(the two last rows in Table I and 7th column of Table II). 

An investigation of
the splitting of the IVDR within this approach into the main peak
which exhausts mainly the model-independent EWSR  and a much broader 
satellite
(pygmy-like resonances with a much smaller contribution to the EWSR
\cite{kievPygmy_voinov,kievPygmy_larsen,nester1,nester2,BMRPhysScr2014} )  
is extended now by taking into account the slope $L$ dependence of the
symmetry energy density. Note that the relative strength of the
satellite mode with respect to that of the main peak 
disappears as $I$ in the limit of symmetric nuclei
\cite{kolmag,kolmagsh}.

\begin{figure*}
\begin{center}
\includegraphics[width=0.7\textwidth]{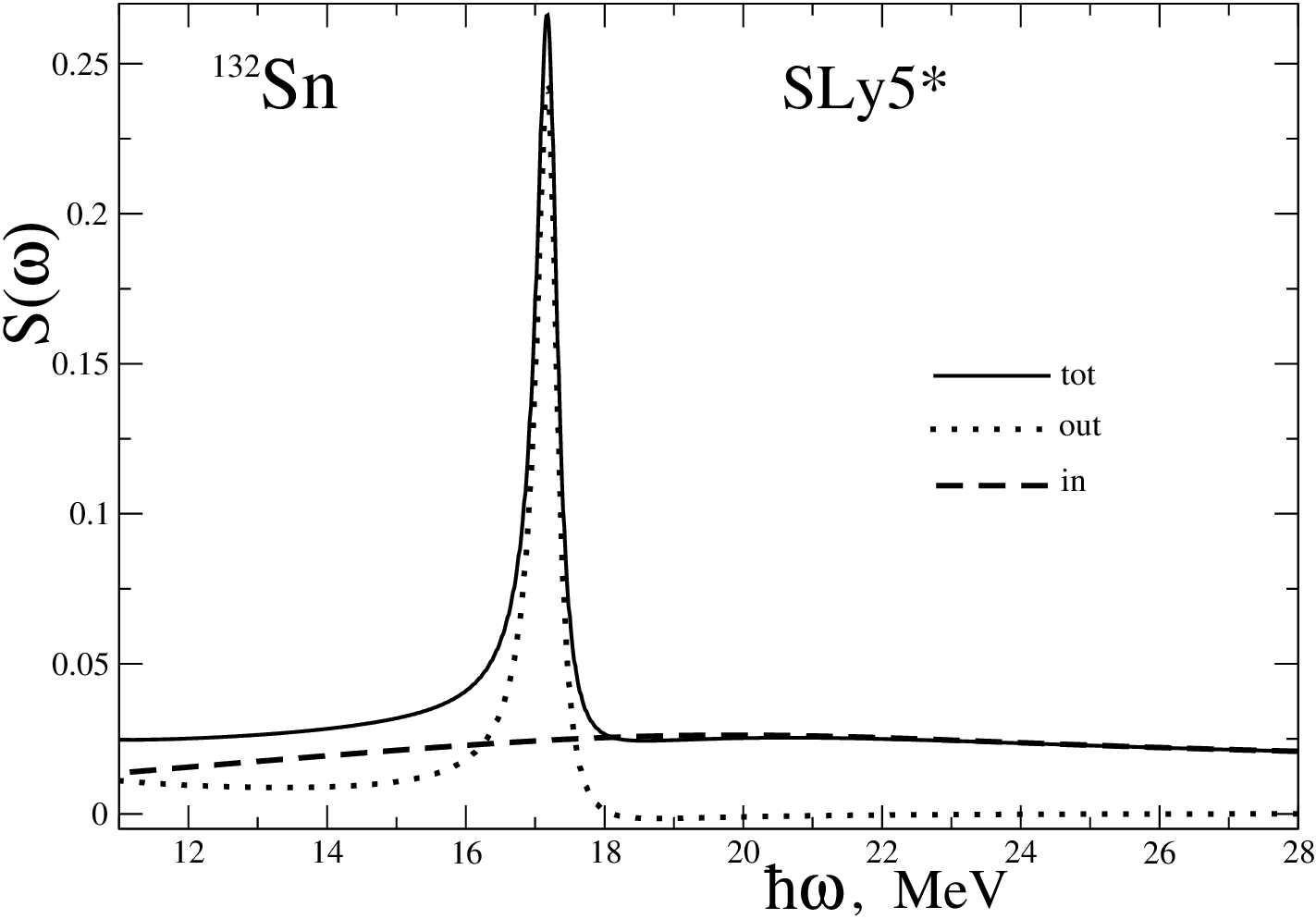}
\end{center}

\vspace{-0.2cm}
\caption{IVDR strength functions 
$S(\omega)$ 
vs the excitation
energy $\hbar \omega$ are shown for vibrations of the nucleus
$^{132}$Sn for the Skyrme force SLy5$^*$ by solid line at $L=50 MeV$;
dotted (``out-of-phase''), and  dashed (``in-phase'') 
curves show separately the main and satellite excitation modes, respectively
\cite{BMRPhysScr2014};
the relaxation time of the collision term 
$T=4.3 \cdot 10^{-21}$ s in agreement with widths of the IVGDRs.} 
\label{fig3}
\end{figure*}

\vspace{0.5cm}
\begin{figure*}
\begin{center}
\includegraphics[width=0.7\textwidth]{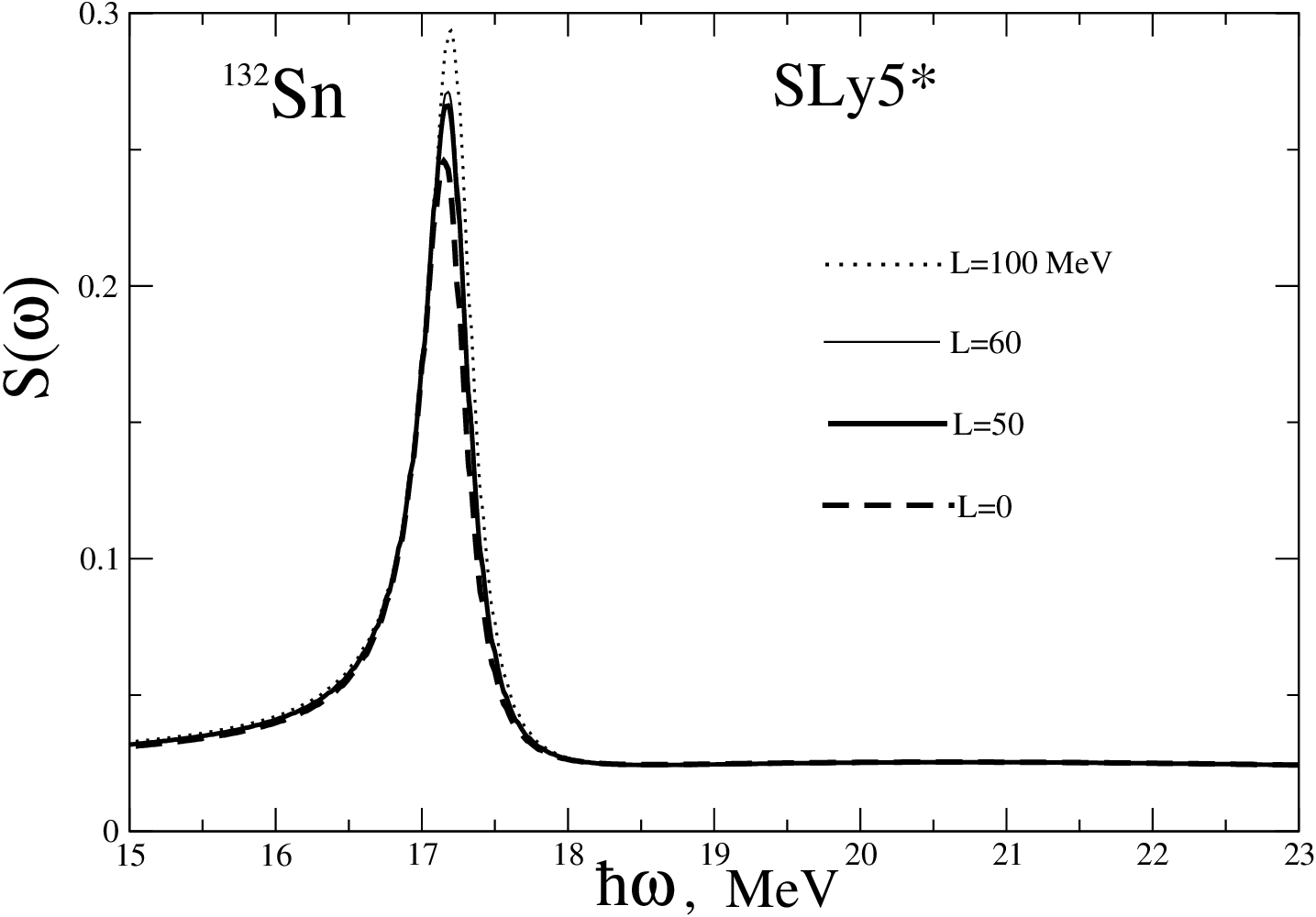}
\end{center}

\vspace{0.1cm}
\caption{The same total strenghts as in Fig.\ \ref{fig3} are shown  
for different $L=0, 50, 60$ and $100$ MeV by dashed, thick and thin solid
and dotted lines. 
} 
\label{fig4}
\end{figure*}
\begin{figure*}
\begin{center}
\includegraphics[width=0.7\textwidth]{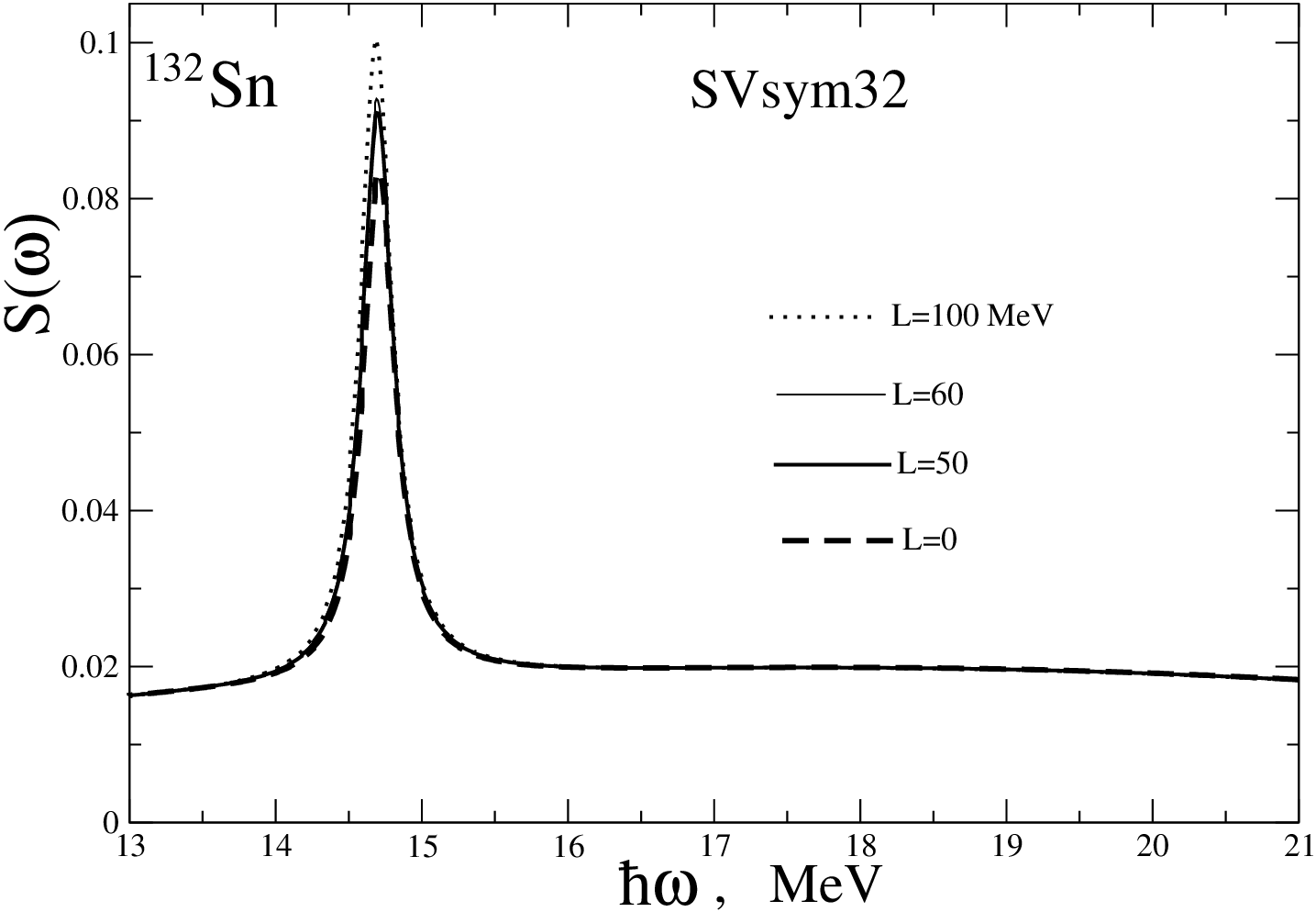}
\end{center}

\vspace{-0.1cm}
\caption{The same as in Fig.\ \ref{fig4} but for SVsym32 Skyrme 
force ($L=60$ MeV) at the relaxation time $T$
with the same constant of the frequency dependence as in  
the previous Figure (\cite{belyaev}),
$T=7.5 \cdot 10^{-21}$ s.}
\label{fig5}
\end{figure*}

The total IVDR strength function being 
the sum of the two (``out-of-phase'' $n=1$ and 
``in-phase'' $n=2$) modes for the isovector- and isoscalar-like 
volume particle density vibrations, respectively 
[Fig.\ \ref{fig3} for the finite $L=50$ MeV] 
have a shape asymmetry \cite{BMRPhysScr2014} (the SLy5$^*$ forces
are taken again as a typical example). The $L$ dependence of the total
IVDR strength is shown
in Fig.\ \ref{fig4} for the SLy5$^*$ and Fig.\ \ref{fig5} 
for SVsym32 cases. The characteristic values of $L$ are used
in these calculations ($L\approx 50$ MeV for
the SLy5$^*$ force \cite{jmeyer} and $L\approx 60$ MeV for
the SVsym32 one  \cite{reinhardSV}).
In Figs.\ \ref{fig3} and \ref{fig4} for the SLy5$^*$ and Fig.\ \ref{fig5} 
for SVsym32 forces, 
one has the ``in-phase'' satellite to the right of the 
main ``out-of-phase''
peak. An enhancement on its left for SLy5$^*$ is due to
the increasing of the ``out-of-phase'' strength (dotted) curve 
at small energies as compared to the IVGDR (Fig.\ \ref{fig3} and \ref{fig4}), 
in contrast to the  
 SVsym32 case shown in Fig.\ \ref{fig5}. 
 The IVDR energies of the two modes in the nucleus $^{132}$Sn do not change
significantly with $L$ in both cases (Table II). 
However, as seen from Table II and Figs.\ \ref{fig4} and \ref{fig5}, 
the L dependence of the EWSRs of these modes for SVsym32
is larger than for SLy5$^*$. Notice,  the essential re-distribution of 
the EWSR contributions among them due to a significant enhancement of 
the main ``out-of-phase'' peak with increasing $L$ for SLy5$^*$ 
(Figs.\ \ref{fig3} and \ref{fig4}) and more pronounced EWSR
re-distribution with $L$ for SVsym32 
(Fig.\ \ref{fig5}) in the same nucleus (Table II).  
Slopes of the $L$ dependence of
$\tau/I$ which is almost linear are approximately the same for both 
considered Skyrmes but smaller  
than those found in \cite{vinas4} probably due to different 
definitions of the neutron skin thickness $\tau$ \cite{BMRV} which are
related to the value of $\delta^{}_S=\rho_{-}/\rho_{+}$ at the nuclear effective 
surface \cite{myswann}.
Note also that these more precise calculations (cf. with  
those of \cite{BMRV,BMRPhysScr2014}) take into account 
higher (4th order) terms of the power expansion 
in the small parameter $\gamma$  for any reasonable 
$L$ change \cite{vinas4}. This is essentially important for the IVDR 
strength distribution for SV forces  because of smaller $c_{\rm sym}$ 
as compared to those for other Skyrme interactions. Constants $\widetilde{c}$  
for the isovector solutions $w_{-}$, (\ref{ysolminus}), are modified 
essentially (besides of the $L$ dependence) by higher order terms due to a 
non-linearity equation for $\psi(w)$ solved in terms of the power series.   
Decreasing the relaxation time $T$ in factor of about 1.5
with respect to the value (Figures) 
evaluated from the data on the 
widths of the IVGDRs at their energies almost does not change the 
IVDR strength structure \cite{belyaev}. However, we found
its strong dependence on $T$ in a more wide value 
region [in factor of about 2-3].
The ``in-phase'' strength component with  
rather a wide maximum is weakly dependent 
on the choice of the Skyrme forces
\cite{chaban_npa,reinhardSV,pastore} and on the slope parameter
$L$, as well as the relaxation time $T$.

The most responsible parameter of the Skyrme HF approach leading to the 
significant differences in the $k_S$ and $Q$ values
 is the constant ${\cal C}_{-}$ (\ref{Cpm}) in 
the gradient terms of the energy density 
(\ref{enerden}). 
Indeed, the key quantity in the expression for $Q$ (\ref{stiffin}) 
and the isovector surface energy constant $k^{}_S$  
(\ref{bsminus}), is the constant ${\cal C}_{-}$
because one mainly has $k^{}_S\propto {\cal C}_{-}$ \cite{BMRV}, 
and $Q \propto 1/k_S \propto 1/{\cal C}_{-}$. Concerning $k^{}_{S}$
and the IVDR strength structure,  this is even more 
important than the $L$ dependence though the latter changes significantly the
isovector stiffness $Q$ and the neutron skin $\tau$. 
The constant ${\cal C}_{-}$ is very different for different Skyrme
forces in the absolute value  and even sign,
 ${\cal C}_{-}\approx -23 $ to $26$ MeV$\cdot$fm$^{5}$ 
($k_S\approx -15$ to $18$ MeV). 
 Contrary to the isoscalar parameters ($b^{(+)}_S$),
there are so far no clear experiments which would determine $k^{}_S$
well enough because the mean energies of the IVGDR (main peaks) do not 
depend very much on $k^{}_S$
for different Skyrme forces (see last row of Table I and 7th column of
Table II).
Perhaps, the low-lying isovector collective states are more sensitive but 
there is no careful systematic study of their
 $k^{}_S$ dependence at the present time.
Another reason for so different $k^{}_S$ and $Q$ values might be traced back
to the difficulties in deducing $k^{}_S$ directly from the HF calculations
due to the curvature and quantum effects in contrast to $b_S^{(+)}$.
It is worthwhile to study the semi-infinite matter within our approach
to avoid such effects and compare with semimicroscopic models.
We have to go also far away from the nuclear stability line to
subtract uniquely the coefficient $k^{}_S$ in the dependence of
$b_S^{(-)} \propto I^2=(N-Z)^2/A^2$, according to (\ref{bsminus}).
For exotic nuclei one has more problems to relate $k^{}_S$ to the
experimental data with a good enough precision.
The $L$ dependence of the neutron skin $\tau$
is essential but not dramatic in the case of SLy and SV
forces (Table I).
The precision of such a description depends more on the specific
nuclear models 
\cite{vinas2}. Our results for the neutron skin  
$R_n-R_p \approx 0.10 - 0.13$ fm in $^{208}$Pb (Table II) in a reasonable 
agreement with the experimental data (Fig.\ 3 of \cite{vinas4}),
$(R_n-R_p)_{\rm exp} = 0.12 - 0.14$ fm (the coefficient
 $(3/5)^{1/2}$ of the square-mean neutron and proton radii was taken 
into account to adopt the definitions, see also
very precised experimental data \cite{osaka}). For $^{132}$Sn our 
analytical evaluation
predict the neutron skin thickness $(R_n-R_p)_{\rm exp} \approx 0.12 - 0.15$ fm
($0.08 - 0.10$ fm for a more known nucleus $^{120}$Sn, also in agreement
with the experimental
data collected in \cite{vinas4}).
The neutron skin thickness $\tau$, like the stiffness $Q$,
is interesting in many aspects for the investigation of exotic nuclei,
in particular, in nuclear astrophysics.

We emphasize that for the specific Skyrme forces there exist an 
abnormal behavior of the
isovector surface constants $k^{}_S$ and $Q$ as related mainly to 
the fundamental constant ${\cal C}_{-}$ of
the energy density (\ref{enerden}) but not much to 
the derivative symmetry-energy density corrections. 
The coefficient $\nu$ (\ref{stiffin}) is almost independent 
on ${\cal C}_{-}$ for SLy and SV Skyrme forces (Table I). 
As compared to 9/4 suggested in \cite{myswann}, 
they are significantly smaller
in magnitude for the most of the Skyrme forces, besides of SkM* 
($\nu \approx 2.3$).

Notice that the isovector gradient terms which are important for
the consistent derivations within the ES approach are 
not included
into the relativistic local density
approaches \cite{vretenar_1,vretenar_2}.
In contrast to all other Skyrme forces, for RATP \cite{chaban_npa} and 
SV \cite{reinhardSV} 
(like for SkI) forces, the isovector stiffness $Q$ is even negative 
as ${\cal C}_{-}>0$ ($k^{}_S>0$),
 that would correspond to the unstable
vibration of the neutron skin.

\section{Conclusions}

Simple expressions for the isovector 
particle densities and 
energies in the leading ES
approximation were obtained analytically, i.e., 
for the surface symmetry energy, the neutron skin thickness
and the isovector stiffness coefficients as functions of the 
slope parameter $L$. We have to include higher order terms
in the parameter $a/R$ to derive the surface symmetry energy.
It depends
on the particle density which can be taken into account  
at leading order in $a/R$. 
These terms depend on the well-known parameters of 
the Skyrme forces. Our results for
the isovector surface energy constant $k^{}_S$,  the neutron skin 
thickness $\tau$ and the stiffness $Q$
depend in a sensitive way on the choice of the parameters of the 
Skyrme energy density (\ref{enerden}), especially on its
gradient terms through
the parameter ${\cal C}_{-}$ (\ref{Cpm}).
Values of the isovector constants $k^{}_S$, $\tau$, and especially, $Q$
 depend also essentially
on the slope parameter $L$, and the spin-orbit interaction constant $\beta$. 
The isovector stiffness constants
$Q$ 
become more close to the empirical data
accounting for their $L$-dependence. 
The mean IVGDR energies and sum rules calculated
in the  
macroscopic models like the FLD  model \cite{kolmagsh,BMRPhysScr2014}
are in a fairly good agreement with the
experimental data. We found a reasonable two-mode main and satellite
structure of the IVDR strength within the FLD model as compared to 
the experimental data and recent other theoretical works.
We may interprete semiclassically the IVDR satellites as 
some kind of the pygmy resonances. Their energies and sum rules 
obtained analytically within the semiclassical FLD approximation 
are sensitive to the surface symmetry energy constant 
$k^{}_S$. Energies $E_1$ and $E_2$ (Table II) are independent of the slope 
parameter $L$ but EWSRs $S_1$ and $S_2$ do depend on $L$, especially for 
SVsym32. It seems helpful to describe them in terms of the only few critical
constants, like $k^{}_S$ and $L$. Therefore, their comparison with the  
experimental data on the IVDR strength splitting  
can be used for the evaluation of $k^{}_S$ and $L$, 
in addition to the experimental data for the neutron skin.

For further perspectives, it would be worth to
apply our results to calculations of pygmy resonances in the IVDR
strength
within the FLD model \cite{kolmagsh} in a more systematic way.
More general problems of the classical and quantum chaos
in terms of the level statistics \cite{BMYnpae2010}
 and Poincare and Lyapunov exponents \cite{BMYijmp2011,BMprc2012}
might lead to a progress 
in studying the fundamental
properties of the collective dynamics like nuclear fission 
within the Swiatecki\& Strutinsky
Macroscopic-Microscopic model. Our approach is helpful also 
for further study of the
effects in the surface symmetry energy because it gives the analytical
universal expressions for the constants $k^{}_S$, $\tau$ and $Q$ 
as functions of the symmetry slope parameter $L$ which are independent 
of the specific properties of the nucleus. 

\begin{ack}
Authors thank 
K.\ Arita, N.\ Dinh Dang, 
M.\ Kowal, V.O.\ Nesterenko, M. Matsuo, K.\ Matsuyanagi, 
J.\ Meyer, T.\ Nakatsukasa, A.\ Pastore, 
P.-G.\ Reinhard, A.I.\ Sanzhur, J.\ Skalski,
and X.\ Vinas for many useful discussions. One of us (A.G.M.) is
also very grateful for a nice hospitality during his working visits at the
National Centre for Nuclear Research in Otwock-Swierk and Warsaw, Poland,
and at the Nagoya Institute of Technology.
This work was partially supported by the Deutsche
Forschungsgemeinschaft Cluster of Excellence 'Origin and
Structure of the Universe' (www.universe-cluster.de) and
Japanese Society of Promotion of Sciences, ID No. S-14130.

\end{ack}


\begin{thebibliography}{99}

\vspace{-0.25cm}
\bibitem{myswann} Myers W D, Swiatecki W J 1969
Ann. Phys. {\bf 55} 395; 1974 {\bf 84} 186
%

\vspace{-0.25cm}
\bibitem{myswnp80pr} Myers W D, Swiatecki W J 1980
Nucl. Phys. {\bf A336} 267

%
\vspace{-0.25cm}
\bibitem{myswiat85} Myers W D, Swiatecki W J and  Wang C S 1985
Nucl. Phys. {\bf A436} 185


\vspace{-0.25cm}
\bibitem{myswnp96pr} Myers W D, Swiatecki W J 1996
Phys. Rev. {\bf C601} 141
%

\vspace{-0.25cm}
\bibitem{vinas1} Centelles M, Roca-Maza X, Vinas X, and Warda M 2009
Phys. Rev. Lett. {\bf 102} 012502 

\vspace{-0.25cm}
\bibitem{vinas2} Warda M, Vinas X, Roca-Maza X and Centelles M 2010
Phys. Rev. C {\bf 82} 054314

\vspace{-0.25cm}
\bibitem{vinas3}  Roca-Maza X,  Centelles M, Vinas X and Warda M 2011
Phys. Rev. Lett. {\bf 106} 252501 

\vspace{-0.25cm}
\bibitem{vinas4} Vinas X, Centelles M, Roca-Maza X and Warda M 2014
Eur. Phys. A {\bf 50} 27

\vspace{-0.25cm}
\bibitem{gross1} Runge E and Gross E K U 1984 Phys. Rev. Lett.
{\bf 52} 997

\vspace{-0.25cm}
\bibitem{gross2} Marques M A L and Gross E K U 2004 Annu. Rev. Chem.,
{\bf 55} 427

\vspace{-0.25cm}
\bibitem{brguehak} Brack M, Guet C, and Hakansson H-B 1985
Phys. Rep. {\bf 123} 275

\vspace{-0.25cm}
\bibitem{brink} Vautherin D and Brink D M 1972 Phys. Rev. C {\bf 5} 626

\vspace{-0.25cm}
\bibitem{bender} Bender M, Heen P-H, and Reinhard P-G 2003 Rev. Mod. Phys. 
{\bf 75} 125
%

\vspace{-0.25cm}
\bibitem{chaban_npa} Chabanat E et al 1997
Nucl.Phys. A{\bf 627} 710; 1998 {\bf A635} 231
%

\vspace{-0.25cm}
\bibitem{reinhardSV} Kl\"upfel P, Reinhard R-G, B\"urvenich T J and
Maruhn J A 2009 Phys. Rev. {\bf C79} 034310
%

\vspace{-0.25cm}
\bibitem{pastore} A. Pastore et al 2013
Phys. Scr. {\bf T154} 014014
%

\vspace{-0.3cm}
\bibitem{strtyap} Strutinsky V M and Tyapin A S 1964 Exp. Theor. Phys. (USSR)
{\bf 18} 664

\vspace{-0.3cm}
\bibitem{strmagbr} Strutinsky V M, Magner A G and Brack M 1984  Z. Phys.,
A {\bf 319} 205 

\vspace{-0.3cm}
\bibitem{strmagden} Strutinsky V M, Magner A G and Denisov V Yu 1985 Z. Phys.
A {\bf 322} 149 

\vspace{-0.3cm}
\bibitem{magsangzh} Magner A G, Sanzhur A I and Gzhebinsky A M 2009
Int. J. Mod. Phys. E {\bf 18} 885

\vspace{-0.3cm}
\bibitem{BMRV} Blocki J P, Magner A G, Ring P and Vlasenko A A 2013
Phys. Rev. {\bf C87} 044304 

\vspace{-0.25cm}
\bibitem{agrawal} Agrawal B K, De J N, Samaddar S K at al 
2013 Phys. Rev. C {\bf 87} 051306

\vspace{-0.25cm}
\bibitem{kievPygmy_voinov} Voinov  A et al 2010 Phys. Rev. {\bf C81} 024319

\vspace{-0.25cm}
\bibitem{kievPygmy_larsen}
Larsen A C et al 2013, Phys. Rev. {\bf C87} 014319

\vspace{-0.25cm}
\bibitem{nester1} Repko A, Reinhard R-G, Nesterenko V O and 
Kvasil J 2013 Phys. Rev. {\bf C87} 024305


\vspace{-0.25cm}
\bibitem{nester2} Kvasil J, Repko A, Nesterenko V O et al 2013
Phys. Scr. {\bf T154} 014019
%

\vspace{-0.25cm}
\bibitem{dang1998} Dinh Dang N, Tanabe K and Arima A 1998 
Phys. Rev. C {\bf 58} 3374; 1999 Phys. Rev. C {\bf 59} 3128
%

\vspace{-0.25cm}
\bibitem{dang2001} Dinh Dang N, Kim Au V, Suzuki T and Arima A 2001 
Phys. Rev. C {\bf 63} 044302
%

\vspace{-0.25cm}
\bibitem{ponomarev} Ryezayeva N, Hartmann T, Kalmykov Y et al 2002
Phys. Rev. Lett. {\bf 89} 272502
%

\vspace{-0.25cm}
\bibitem{adrich} Adrich P et al 2005 Phys. Rev. Lett. {\bf 95} 132501  
%

\vspace{-0.25cm}
\bibitem{wieland} Wieland O  et al 2009  Phys. Rev. Lett. 
  {\bf 102} 092502
%

\vspace{-0.25cm}
\bibitem{nakatsukasa2012}Tohyama M and Nakatsukasa T 2012 Phys. Rev. C {\bf 85}
031302(R)

\vspace{-0.25cm}
\bibitem{BMRPhysScr2014} Blocki J P, Magner A G and Ring P 2014
Phys. Scr. {\bf T89} 054019


\vspace{-0.25cm}
\bibitem{kolmag} Kolomietz V M and Magner A G 2000 Phys. Atom. Nucl. {\bf 63}
1732 

\vspace{-0.25cm}
\bibitem{kolmagsh} Kolomietz V M, Magner A G, and Shlomo S 2006
Phys. Rev. {\bf C73} 024312
%

\vspace{-0.25cm}
\bibitem{BMR2015}  Blocki J P, Magner A G and Ring P 2015 to be published.
%

\vspace{-0.25cm}
\bibitem{jmeyer}  Meyer J 2014, private communications.
%

\vspace{-0.25cm}
%
\bibitem{vretenar_1} Vretenar D, Paar N, Ring P and Lalazissis G A 2001
Phys. Rev. {\bf C63} R13
%

\vspace{-0.25cm}
\bibitem{vretenar_2}
 Vretenar D, Paar N, Ring P and Lalazissis  G A 2001
Nucl. Phys. {\bf A692} 496
%

\vspace{-0.25cm}
\bibitem{belyaev} Magner A.G, Gorpinchenko D V and Bartel J 2014
Phys. At. Nucl. {\bf 77} 1229 

\vspace{-0.25cm}
\bibitem{osaka} Tamii et al 2011 Phys. Rev. Lett. {\bf 107} 062502

\bibitem{BMYnpae2010} Blocki J P, Magner A G and Yatsyshyn I S 2010 
At. Nucl. Energy {\bf 11} 239

\bibitem{BMYijmp2011} Blocki J P, Magner A G and Yatsyshyn I S 2011 
Int. J. Mod. Phys. E 20, 292 

\bibitem{BMprc2012} Blocki J P and Magner A G 2012 Phys. Rev. C {\bf 85} 064311 


\end{thebibliography}
\end{document}